\documentclass{PoS}

\usepackage{graphicx}
\usepackage{grffile}
\usepackage{amsmath}
\usepackage[utf8]{inputenc}

\title{Supermultiplets of the $\mathcal{N} = 1$ supersymmetric Yang-Mills 
theory in the continuum limit}

\ShortTitle{Supermultiplets of the $\mathcal{N} = 1$ supersymmetric 
Yang-Mills theory in the continuum limit}

\author{\speaker{Pietro Giudice}, Gernot M\"unster\\
  Universit\"at M\"unster, Institut f\"ur Theoretische Physik, \\
  Wilhelm-Klemm-Str. 9, D-48149 M\"unster, Germany\\
  E-mail: \email{p.giudice@uni-muenster.de, munsteg@uni-muenster.de}}

\author{Georg Bergner\\
  Universit\"at Bern, Institut f\"ur Theoretische Physik, \\
  Sidlerstr.~5, CH-3012 Bern, Switzerland\\
  E-mail: \email{bergner@itp.unibe.ch}
}

\author{Istvan Montvay\\
  Deutsches Elektronen-Synchrotron DESY, \\
  Notkestr. 85, D-22603 Hamburg, Germany\\
  E-mail: \email{montvay@mail.desy.de}
}

\author{Stefano Piemonte\\
Universität Regensburg, Institute for Theoretical Physics, \\
D-93040 Regensburg, Germany \\
  E-mail: \email{stefano.piemonte@ur.de}
}

\abstract{
The spectrum of $\mathcal{N} = 1$ supersymmetric Yang-Mills theory, 
calculated on the 
lattice, is presented. The masses have been determined on three different 
lattice spacings and extrapolated towards vanishing gluino mass. We present 
the extrapolation to the continuum limit which is consistent with the 
formation of degenerate supermultiplets.}

\FullConference{The 33rd International Symposium on Lattice Field Theory\\
		 14 -18 July  2015\\
		 Kobe International Conference Center, Kobe, Japan}

\newcommand{\aetap}{\text{a--}\eta'}
\newcommand{\api}{{\text{a--}\pi}}
\newcommand{\afn}{\text{a--}f_0}

\newcommand{\tr}[1]{\ensuremath{\mathrm{Tr}} \left[{#1}\right]}
\newcommand{\I}{\ensuremath{\mathrm{i}\hspace{1pt}}}

\newcommand{\beq}{\begin{equation}}
\newcommand{\eeq}{\end{equation}}
\newcommand{\bea}{\begin{eqnarray}}
\newcommand{\eea}{\end{eqnarray}}

\begin{document}

\section{Introduction}

The Standard Model (SM) of particle physics works pretty well.
Still there is what is perceived by many theorists as an
imperfection of the theory: that is the so-called fine-tuning problem.
The problem arises because the mass of the Higgs boson is much lighter
than the Planck mass, {\it i.e.} the scale when quantum gravity becomes 
important. 
Because of this, large quantum contributions would make the mass 
of the Higgs huge, unless an incredible fine-tuning cancellation 
occurs.
Supersymmetry (SUSY), introducing for every SM particle a superpartner, 
protects the mass of the Higgs eliminating the quadratic
mass divergences responsible for the problem.

Assuming that the standard model of cosmology is correct, the
Planck mission team concluded that the total mass/energy density of the known 
universe contains $4.9\%$ of ordinary matter, $26.8\%$ of dark matter  
and $68.3\%$ of dark energy~\cite{Ade:2015xua}.
The main hypothetical particle candidates for dark matter are
known as weakly interacting massive particles, WIMPs. 
In some supersymmetric models the lightest supersymmetric particle
is stable and electrically neutral. Moreover, it interacts weakly with the 
particles of the Standard Model, becoming therefore a good candidate 
for dark matter.

SUSY is therefore a prominent candidate for an extension of the 
Standard Model, but unfortunately so far, after Run 1 of LHC, no direct 
evidence for SUSY has been found~\cite{Bechtle:2015nta, sfyrla:2015}.

It is worth noting that actually the first response  to the fine-tuning problem
was to propose that the symmetry breaking occurs {\it dynamically}. 
Such theories
are generically called technicolor. Our collaboration has recently started
to work on one of them~\cite{Bergner:lat2015}.

Until a few years ago, the effects of SUSY on a theory have been studied only
by means of perturbative or effective methods. However, SUSY can be studied 
also non-perturbatively using the lattice field theory approach,
{\it i.e.} discretising the space-time onto a lattice and then 
performing Monte Carlo simulations. 

Because SUSY is explicitly broken on the lattice, as a consequence of the 
absence of translational invariance, different approaches are
considered in literature.
Some groups, {\it e.g.}~\cite{Asaka:2013tye, Catterall:2015ira}, work on 
formulating theories which have exact lattice supersymmetry on the lattice.
The approach of our collaboration is based on the idea that, by appropriate
tuning of the parameters of the theory, SUSY has to be recovered in the 
continuum limit. In these proceedings we show that this program can be
pursued. We have determined the mass spectrum of the theory, before 
extrapolating it to the chiral limit, and than to the continuum limit.
The formation of degenerate supermultiplets, testifying the restoration
of SUSY, is observed.
This contradicts a spontaneous supersymmetry breaking, in principle possible, 
in the supersymmetric Yang-Mills (SYM) theory we are considering.

\section{The properties of the theory}
The theory we study is $\mathcal{N} = 1$ SYM theory
with gauge group SU(2). The theory contains gluons as bosonic particles 
and gluinos as their fermionic superpartners.
The Lagrangian of the theory is invariant under a transformation which 
relates gluons, $A_\mu(x)$ gauge fields, and gluinos, $\lambda(x)$ fields:
\bea
A_\mu(x) & \rightarrow & A_\mu(x) -2 \,\I \bar{\lambda}(x)\gamma_\mu \epsilon \\
\lambda^a(x) & \rightarrow & \lambda^a(x) - \sigma_{\mu\nu} F^a_{\mu\nu}(x) 
\epsilon \ ,
\eea
where $\epsilon$ is a global fermionic parameter (consistent with the 
Majorana condition), parametrising the transformation.

The fundamental assumption about non-perturbative dynamics of SYM theory 
is that there is confinement of the fundamental colour charge and 
spontaneous chiral symmetry breaking. Both properties have been 
observed in our simulations~\cite{Montvay:2001aj}.

In our theory there is only one Majorana adjoint flavour, therefore there
is only a global abelian chiral symmetry U(1)$_\lambda$. This classical
symmetry is broken at the quantum level, {\it i.e.} it is anomalous. 
The remnant symmetry
is a $Z_4$ one; this discrete symmetry is spontaneously broken to $Z_2$, 
as can be revealed by the condensation of the gluino 
$\langle \lambda \lambda \rangle$.

Confinement manifests itself as a linear increase of the potential between 
two static sources. The consequence is that the spectrum of the theory consists 
of colourless hadronic bound states. In the SUSY limit, the masses
of particles are organised in degenerate multiplets. 

The low-lying spectrum of particles has been predicted constraining the form 
of the low-energy effective actions by means of the symmetries of the theory.
In the seminal work~\cite{Veneziano:1982ah} a first supermultiplet was 
described:
it consists of a scalar ($0^+$ gluinoball: $\afn \sim \bar\lambda \lambda$),
a pseudoscalar ($0^-$ gluinoball: $\aetap \sim \bar\lambda \gamma_5 \lambda$),
and a Majorana fermion (spin $1/2$ gluino-glueball:
$\chi \sim \sigma^{\mu \nu} \tr{ F_{\mu \nu} \lambda }  $).
A second supermultiplet was found in~\cite{Farrar:1997fn} introducing 
pure gluonic states in the effective Lagrangian. It consists of
a $0^-$ glueball, a $0^+$ glueball, and a gluino-glueball.
In that paper the authors argued that the glueball states, in 
the supersymmetric limit, are lighter than the gluinoball states.
Other authors~\cite{Feo:2004mr}, using different arguments, and hints from 
ordinary QCD, deduce that the lighter states are instead gluinoballs. 
Our work is able to shed new light on this problem as well.

\section{The theory on the lattice}
The formulation we employ in our simulations is an improved version
of what was first proposed in~\cite{Curci:1986sm} where the Wilson-Dirac 
operator was considered. For the gauge part, we use the tree-level Symanzik 
improved gauge action including rectangular Wilson loops of 
perimeter's~length six.
To reduce the lattice artifacts also in the fermionic part we apply one
level of stout smearing~\cite{Morningstar:2003gk} to the link fields in the 
Wilson-Dirac operator. The configurations have been obtained by updating
with a two step polynomial hybrid Monte Carlo (TS-PHMC) 
algorithm~\cite{Montvay:2005tj}. The integration of the fermionic variables
yields a Pfaffian: the sign is taken into account by 
reweighting~\cite{Demmouche:2010sf}.

On the lattice there are three sources of SUSY breaking:
1) A non-zero gluino mass; it has to be introduced in numerical simulations 
but it breaks supersymmetry softly (is does not spoil the cancellation of 
quadratic divergences which SUSY is supposed to cure). 
2) Finite volume effects; in~\cite{Bergner:2012rv} we have shown that using 
a box size of about $1.2$~fm (in QCD units) the statistical errors and the 
systematic errors of the finite size effects, with our current numerical 
precision, are of the same order and hence the finite size effects can be 
neglected. 
3) The discretisation of the space-time on a lattice and the consequent 
breaking of the translational invariance~\cite{Bergner:2009vg}.
Because the generators of translation $P_\mu$ appear
directly in the relation which describes the algebra of the 
single conserved supercharge $Q$,
\beq
\{ Q_\alpha,Q_\beta \} \propto P_\mu \  \quad 
\mbox{($Q_\alpha$ are Majorana spinors)}  \ ,
\label{commutationrelation}
\eeq
than because the symmetry associated with the r.h.s is broken, also the one 
on the l.h.s, {\it i.e.} SUSY, is broken.

As discussed in~\cite{Bergner:2009vg} a way to circumvent this last problem
and to realise complete supersymmetry on the lattice is to introduce
in the Lagrangian nonlocal interactions and a nonlocal derivative. 
The approach followed in our work is different: the idea is that in
the continuum limit Eq.~(\ref{commutationrelation}) is valid again.
Moreover, as shown in~\cite{Curci:1986sm,Suzuki:2012pc}
in SYM the fine-tuning of a single parameter is enough to recover not only 
supersymmetry but also to cancel the explicit chiral symmetry breaking 
induced by Wilson fermions.
In practice, we have to tune the bare gluino mass so that 
the renormalised gluino mass vanishes. This has been verified in the 
past~\cite{Farchioni:2001wx} using the Ward Identities. Currently,
we  monitor the mass of the unphysical adjoint pion $\api$, which is 
defined by the connected contribution of the $\aetap$ correlator.
Using the OZI approximation it was possible to show~\cite{Veneziano:1982ah}
that the square of the adjoint pion mass is proportional
to the mass of the gluino: $m^2_\api \propto m_g$. This relation has
been obtained again recently in a partially quenched 
setup~\cite{Munster:2014cja}.

One of the main issues in the lattice approach is the setting of the scale:
it is fundamental both for obtaining the observables in lattice units and for
extrapolating to the continuum limit the numerical results.
In the last years mainly two different observables have been considered: 
the Sommer parameter $r_0$~\cite{Sommer:1993ce}  and the Wilson flow 
scale $w_0$~\cite{Borsanyi:2012zs}. 
In this work we have considered both these scales and studied them 
systematically~\cite{Bergner:2014ska}.

\section{Mass spectrum numerical results}
The main aim of our simulations is to verify the existence of a continuum 
limit with unbroken supersymmetry. This task is pursued measuring
the low-lying mass spectrum of bound states: we expect in fact degeneracy 
inside the same supermultiplet.

We  have already published results on the spectrum of the theory for 
two values of the lattice spacing. 
Simulations performed at  $\beta=1.60$, corresponding to a lattice spacing 
in QCD units of~$\sim~0.086$~fm, have been presented 
in~\cite{Demmouche:2010sf}.
In \cite{Bergner:2013nwa,Bergner:2013jia} we decreased the value of
the lattice spacing by~$\sim~40\%$ using $\beta=1.75$.
In~\cite{Bergner:2014iea} we presented our preliminary results at $\beta=1.90$,
where the lattice spacing was further reduced by $\sim 30\%$, {\it i.e.}
$a \sim 0.036$~fm. Here we present our final results at $\beta=1.90$, with 
the extrapolation to the continuum limit of the spectrum.
\begin{figure}[t]
\vspace{-6mm}
\phantom{.}
\hspace{-5mm}
\includegraphics[width=7.9cm]{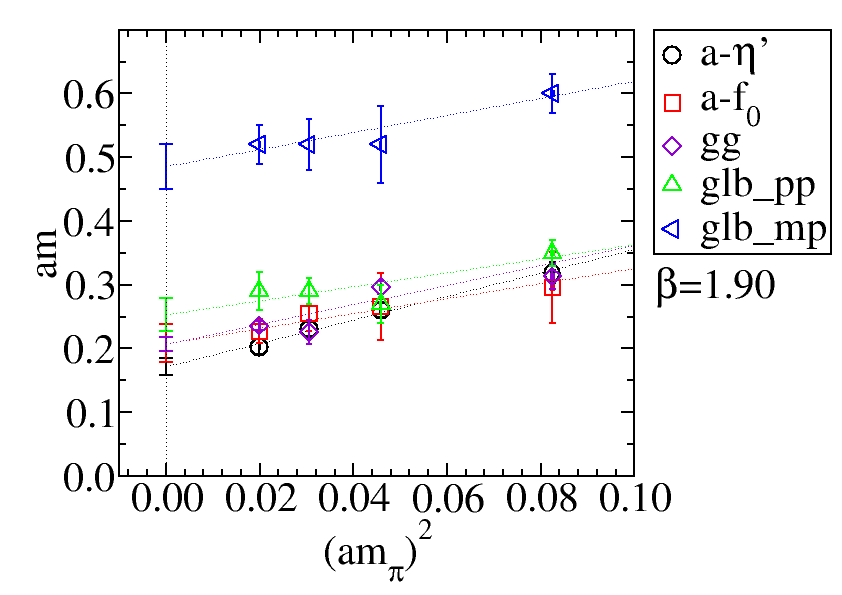}
\includegraphics[width=7.9cm]{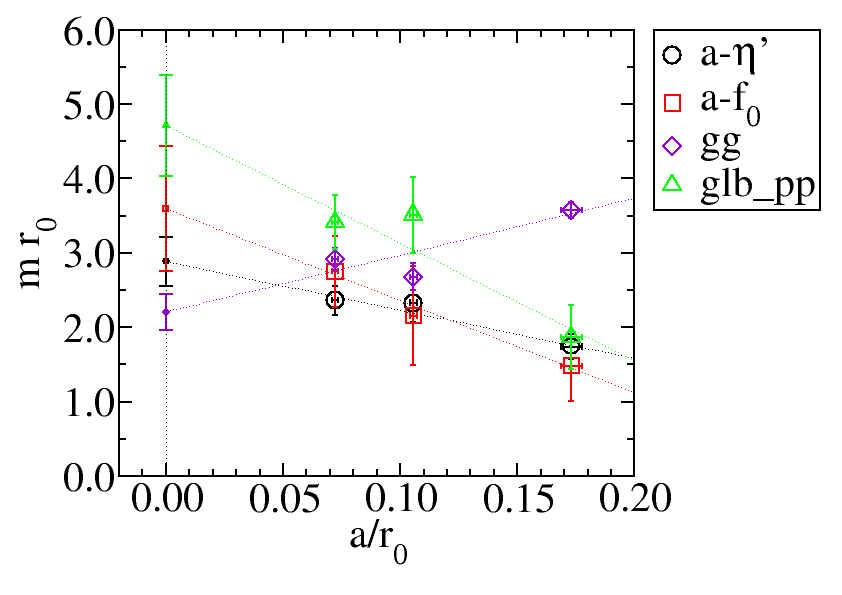}
\phantom{.}
\vspace{-7mm}
\caption{\textbf{(Left)} Mass spectrum (gg: gluino-glue; glb$\_$pp: glueball $0^{++}$; 
  glb$\_$mp: glueball $0^{-+}$) at $\beta=1.90$ for four values of 
  the $\api$ mass. The extrapolated chiral limit is shown as well.
  \textbf{(Right)} Mass spectrum extrapolated to the continuum limit, using $r_0$.}
\label{fig:spectrum}
\end{figure}
In Figure~\ref{fig:spectrum} (Left) we plot five
bounds states for different values of the adjoint pion mass squared;
they are linearly extrapolated to the chiral limit, according to 
the relation $m^2_\api \propto m_g$.
This numerical extrapolation, at fixed lattice spacing, shows already 
degeneracy of the masses in the lowest supermultiplet within two 
standard deviations. 
Note that the glueball $0^{++}$ and the gluinoball $\afn$ have the 
same quantum numbers and they can be characterised by a strong
mixing. The same is true for the glueball $0^{-+}$ and the gluinoball $\aetap$.
This means that it is actually difficult to say which of
the two supermultiplets is more glueball-like or 
gluino-like~\cite{Farrar:1997fn,Feo:2004mr}.
Moreover, to make things more complicated, the operators we 
have chosen to describe the two states can have a strong or
a weak projection with one or the other state.

Now if two states (with two different energies) have a strong
mixing, it seems reasonable that also the operators which describe them will
have a strong overlap with the two states, and extracting the 
masses using these operators we will get the same mass from both of them, 
namely the one of the lightest state.
Vice versa, if the mixing of the states is weak, it is possible that the 
two operators will project differently in the two states, and extracting 
the mass from them we will have two masses with different values.

Looking at Figure~\ref{fig:spectrum} (Left) we can therefore say that
the $0^{++}$ and the $\afn$  have a strong mixing and therefore  
a good overlap with the ground state;
the $0^{-+}$ and the $\aetap$ have a weak mixing and the operators
we are using for the $0^{-+}$ have a strong overlap with the 
glueball-like state, but very weak with the gluino-like.
This line of reasoning seem to indicate that the glueball states have
an energy higher than the gluinoball as argued in~\cite{Feo:2004mr}.

In Figure~\ref{fig:spectrum} (Right), we show the spectrum of the 
theory extrapolated to the continuum limit using the inverse of the 
Sommer scale $r_0$. 
According to this plot we can finally claim that the theory shows
restoration, within two standard deviations, of supersymmetry in 
the continuum limit. A similar plot has been obtained also using
the scale $w_0$, showing an even better behaviour in the
extrapolated continuum limit (even if also in that case we see
degeneracy only within two standard deviations).

\section{Influence of topology on the scale setting}
In a recent paper~\cite{Bergner:2014ska} we have investigated
the relation between the scale, defined by $w_0$ and $r_0$, and the 
topological charge, verifying a correlation between these two quantities.
In this contribution we consider only the Sommer parameter case.

\begin{figure}[t]
\vspace{-9mm}
\phantom{.}
\hspace{-5mm}
\includegraphics[width=7.9cm]{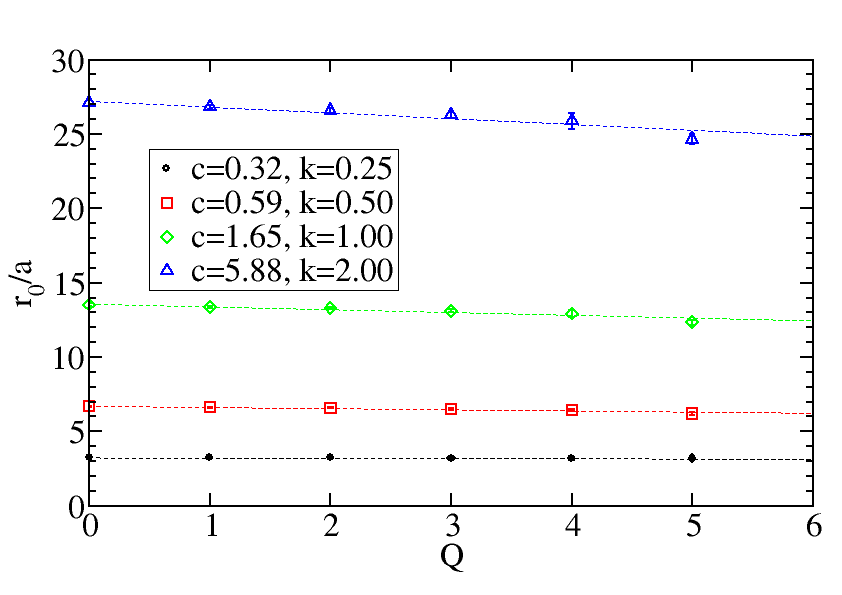}
\includegraphics[width=7.9cm]{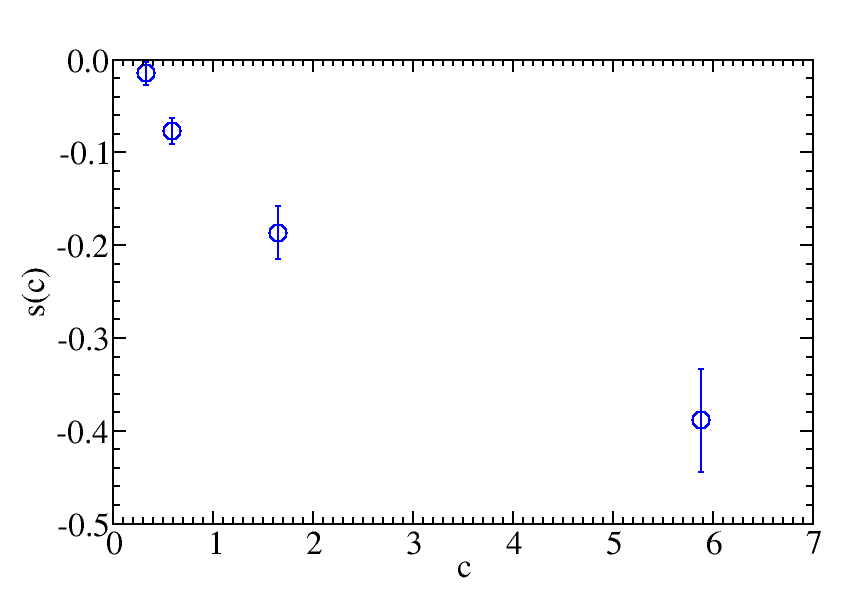}
\vspace{-4mm}
\caption{\textbf{(Left)} Linear fit of the dependence of $r_0$ on the 
topological charge for different values of the parameter $c$. 
The value of the parameter $k$ is the ratio between the interquark distance
(quark-antiquark) for that value of $c$ with respect the standard case $c=1.65$.
\textbf{(Right)} Slope coefficient $s(c)$ as a function of the reference 
value $c$. Lattice $32^3\times 64$, $\beta=1.90$, $\kappa=0.14415$.}
\label{fig:slope}
\end{figure}

$r_0$ is defined as the distance $r$ such that the strong force between a 
static quark-antiquark pair, multiplied by the squared distance,
is equal to a specific reference value, typically 1.65. 
For a generic reference value $c$ we have:
\beq
r_0^2 \left( \frac{\partial V}{\partial r} \right)_{r_0} = c \ .
\eeq
In Figure~\ref{fig:slope} (Left) we have plotted the value of the scale
parameter $r_0$, for different values of the reference parameter $c$,
as a function of the topological charge $Q$. 
APE smearing has been used as smoothing procedure to suppress the short 
distance fluctuations of the lattice topological charge.

Data with the same value of $c$ can be interpolated very well 
($\chi^2/\mbox{d.o.f.} \ \lesssim 1$)  with a linear
fit. The slope $s(c)$ as a function of the reference parameter $c$ 
is plotted in Figure~\ref{fig:slope} (Right). It is evident that
increasing the value of $c$ corresponds to increase also 
the value of the slope. This has important consequences in the 
precision we can determine the scale $r_0$: 
if the reference value is chosen too large, because of the
slope which is large, the fluctuation in the value of $r_0$, 
going from one configuration to 
another, will be large (during the Monte Carlo simulation all 
topological sectors will be explored). 
The error which will characterise the scale parameter will be large. 
Of course the value of $c$ cannot be taken too small otherwise the
discretisation effects will strongly effect the scale determination.

The lesson we learnt from this analysis is that the value of the reference
scale has to be chosen very carefully: there exist an optimal value
which is at the same moment  not affected by the discretisation effects
and which is characterised by the smallest possible value of its statistical 
error. In~\cite{Bergner:2014ska} we have shown that this conclusion is true
also for the Wilson parameter $w_0$.

\section{Conclusion and Outlooks}
In this work we have studied $\mathcal{N} = 1$ supersymmetric Yang-Mills 
theory with gauge group SU(2). We have determined the spectrum of
the theory extrapolating the results first to the chiral limit and
then to the continuum limit. 
We observe approximate degeneracy, within our systematic and statistical 
errors, of the mass spectrum in the low-lying 
supermultiplet indicating the recovery of SUSY in the continuum limit.
In~\cite{Bergner:2015cqa} we have shown that it is possible
to check the SUSY realisation looking directly at the Witten index.
Moreover we have studied the influence of topology on the scale setting,
showing a correlation between the precision with which we can determine 
the scale
and the reference value used in setting the scale.
The next steps of this project include a study of the excited states and 
a deeper analysis of the mixing between glueball-like and gluino-like states. 
Moreover, we will soon move from SU(2) to the more realistic SU(3) gauge theory
which is another important part of the supersymmetric extension of the 
Standard Model.

\section*{Acknowledgements}
This project has been supported by the German Science Foundation (DFG) under
contract Mu 757/16.
The authors gratefully acknowledge the computing time granted by the John
von Neumann Institute for Computing (NIC) and provided on the supercomputers
JUQUEEN and
JUROPA at J\"ulich Supercomputing Centre (JSC).
Further computing time has been
provided by the compute cluster PALMA of the University of M\"unster.


\end{document}